\documentclass[twocolumn,times]{aastex631}
\usepackage{xfrac}
\usepackage{hyperref}
\usepackage{xcolor}  % Required for color definitions
\usepackage[dvipsnames]{xcolor}
\definecolor{xlinkcolor}{cmyk}{1,1,0,0}

\hypersetup{linkcolor=xlinkcolor,citecolor=blue,filecolor=cyan,urlcolor=blue}

\begin{document}

\title{Age Discrepancy in Three Galactic Cepheid Binaries}

\author[0009-0002-2739-070X]{Sashwat Prasadh}
\affiliation{Department of Physics \& Physical Oceanography, Memorial University of Newfoundland \& Labrador, St. John’s, NL A1C 5S7, Canada}
\correspondingauthor{Sashwat Prasadh; \href{mailto:sprasadh@mun.ca}{sprasadh@mun.ca}}

\author[0000-0002-7322-7236]{Hilding R. Neilson}
\affiliation{Department of Physics \& Physical Oceanography, Memorial University of Newfoundland \& Labrador, St. John’s, NL A1C 5S7, Canada}

\author[0000-0002-4374-075X]{Nancy Remage Evans}
\affiliation{Centre for Astrophysics | Harvard \& Smithsonian Astrophysical Observatory, MS 4, 60 Garden St. Cambridge, MA 02138, USA}  

\begin{abstract}
SV Per, RW Cam, and KN Cen are Galactic classical Cepheids found in binary systems along with B-type companions. Each system exhibits an ``apparent age discrepancy", where the Cepheid appears significantly younger than its companion. We compute stellar evolution tracks using the MESA code and compare them with these stars on the Hertzsprung-Russell diagram to estimate the masses and the corresponding ages. We find that, for two of the systems,  the companions appear to be between 1.5 $\times$ to double the age of the Cepheids, suggesting that the Cepheids may be rejuvenated merger products of two main-sequence stars that have since evolved. We find that the third system KN Cen is consistent with single star evolution when we reevaluate ultraviolet observations and find that the companion is hotter than previously measured. We test this idea using the MESA code for each system by evolving a single star model and rapidly accreting mass on the main sequence to emulate a merger, and then continuing the evolution until the model is consistent with the parameters of the Cepheids. The results are consistent with the hypothesis that the Cepheids were originally merger products. We conclude with a  discussion of potential observational tests of the merger hypothesis.
\end{abstract}


\section{Introduction} \label{sec:intro}
Classical Cepheids are intermediate-mass supergiant stars that pulsate radially and reside within the Cepheid instability strip \citep{Turner.2001OAP....14..166T} of the Hertzsprung-Russell Diagram (HRD). These stars have shaped astrophysics and cosmology research for the past century, cementing Cepheids as stellar evolution laboratories owing to the relationship between the pulsation period and luminosity, also known as the Leavitt Law (hereafter PL relation) \citep{Leavitt.Pickering.1912HarCi.173....1L}. The PL relation is the primary standard for measuring distances beyond the scope of parallax \citep{Morel.2010A&A...520A..41M} and has been at the forefront for measurements to distant galaxies. The Gaia mission is revolutionizing the calibration of the PL relation \citep{Clementini.2024IAUS..376..115C} and has significantly contributed to our understanding of Cepheids. Nonetheless, several questions regarding their intrinsic properties remain unresolved.

One of the prime challenges for understanding Cepheids has been the disagreement between the stellar evolution and stellar pulsation mass estimates, termed the ``Cepheid mass discrepancy" \citep{Cox.1980tsp..book.....C}. \citet{Bono.2006MmSAI..77..207B} proposed resolutions in the forms of radiative opacity changes in the envelope \citep{Guzik.2021ASPC..529...79G}, pulsation driven mass loss \citep{Neilson.2008ApJ...684..569N}, and mixing either due to convective core overshooting \citep{Vitense.1958ZA.....46..108B} or resulting from rotation \citep{Anderson2015IAUS..307..206A}. \citet{Miller.2020ApJ...896..128M} demonstrated that most of these theories remain plausible explanations for various Cepheid phenomena. Significant progress has been made over the past few decades to reduce the discrepancy to 20\% or less \citep{Guzik.2023arXiv230712386G}. Even so, luminosity estimates from observations are still up to 30\% higher than stellar evolution tracks \citep{Guzik.2021ASPC..529...79G}. In essence, a portion of this study implies that mass discrepancy and PL relation calibrations continue to affect the agreement between Cepheids masses as described in due course.

 Three recent mass measurements for Cepheids SU Cyg, V1334 Cyg, and Polaris incorporating interferometry as summarized in \cite{2025A&A...693A.111G}.  As
illustrated in their Fig. 10, the masses are increasingly precise, but all
three masses are smaller than those best fit by evolutionary models.  
From observations, about 60 to 80\% of known Galactic Cepheids are in binary systems \citep[][and references therein]{Rathour.2024arXiv240314039S}. However,  there are no known eclipsing binaries in the Milky Way \citep{Udalski.2015AcA....65..341U}.
As such, the measured properties of the companion stars in a single line spectroscopic binary system have a typical accuracy of only 10 to 20\% \citep{Pilecki.2024.II10-20.arXiv240312390P,Evans.2018ApJ...866...30E}. Among the substantial proportion of Milky Way Cepheids anticipated to exist within binary systems, only 60 of them exhibit spectroscopically confirmed companions \citep{Karczmarek.2023ApJ...950..182K,Pilecki.2025arXiv250109793P}.

The unaccounted presence of companions raises questions about the dispersion of the PL relation, considering that these companions can lead to greater observed luminosities compared to lone Cepheids with the same period \citep{Pilecki.2024IAUS..376..150P}.  

Cepheid progenitors are main sequence B stars.  Since B stars are frequently
found in binary or multiple star systems, it is highly likely that some close
B star systems will undergo Roche lobe overflow as they evolve along the main sequence.  Some will probably coalesce into a more massive star which may
ultimately evolve to become a Cepheid.  Identifying merger products is
a challenge except through contexts such as blue stragglers in clusters.
the purpose of this study is to investigate another means of identifying
merger products. 

Studies of binary Cepheid systems with the {\it International Ultraviolet
Explorer (IUE)} satellite have produced new insights for systems with a
hot companion. In these cases the companion dominates the spectrum
at ultraviolet wavelengths, allowing an effective temperature to de determined for the companion.  These
companions are typically main sequence stars.  However, as discovered by
\citet{Vitense1984IAUS..105..449B}, the members of some systems do not match
the same isochrone as the Cepheid.
In a discussion of eight systems with companions not on
the main sequence, \cite{Evans.1994ApJ...436..273E} found that for the systems SV Per, RW Cam and
KN Cen the companions are too cool to be consistent with an isochrone
which matches the Cepheid.

%The merger scenario was suggested for Polaris to explain the apparent age
%difference between Polaris A (the Cepheid) and Polaris B \citep{Bond.2018ApJ...853...55B}.

In this paper, we consider three such fundamental-mode Galactic Cepheids, namely \object{SV Per}, \object{RW Cam}, and \object{KN Cen}, each found in non-eclipsing single line spectroscopic binary systems with less massive B-type companions. These three binary Cepheid systems appear to have an apparent age discrepancy between the Cepheid and its companion as found by \citet{Vitense1984IAUS..105..449B} and \citet{Evans.1994ApJ...436..273E}. We investigate the age discrepancy in SV Per, RW Cam, and KN Cen and hypothesize that the Cepheids evolve from a stellar merger that occurred during main sequence evolution. The structure of the paper is as follows: in the next section we revisit the analysis of the potential the three Cepheids that we hypothesize could be merger products, section \ref{sec:mass} is dedicated to computing the mass of the stars, and section \ref{sec: accretion} describes the assumptions and structure of the merger model that we construct. 
Section \ref{Sec: Discussion} outlines potential resolutions to the age discrepancy and examines observational signatures indicative of mergers in Cepheids.

\section{Possible Merger Targets Revisited}

In this section we reexamine the ultraviolet data from which the
temperature of the companion is derived.

\subsection{KN Cen} 
Of the three Cepheids suggested here as candidates for post-merger products,
KN Cen has the hottest companion.  It is also of  interest because
the Cepheid has a particularly long pulsation period (34.03 d).

The previous discussion was based on a weak IUE spectrum SWP10081 (Exposure
time 420 m, with an exposure level of 67 data numbers out of an ideal 200 DN).
Since the discussion of \cite{Evans.1994ApJ...436..273E}, a stronger exposure has been obtained
SWP 48386 (exposure time 740 min 103 DN).
A summed spectrum has been created from the two longest exposures of
KN Cen B.  

Additional input into the discussion of the temperature of the companion comes
from the photometry of  \cite{1985SAAOC...9....5C} which can be used to
determine the reddening of the system.    Because KN Cen is a long
period Cepheid, it is bright and the contribution of the companion to the
colors has a comparatively small effect.

Determination of the temperatures of the companions from the ultraviolet
spectra, in general, is made from comparison with the 
grid of model atmospheres from \cite{2017AJ....153..234B}.  However, they caution
that for temperatures hotter than approximately 15,000 K, the lack of NLTE
effects distorts the comparison.  This affects the spectra of
KN Cen B but not RW Cam B or SV Per B.
A second factor in the analysis of the spectrum of KN Cen B, the companion,
is that for early B stars energy distribution in
the region between 1200 and 2000 \AA is much
less temperature sensitive than it is for late B stars.  The flux for
early B stars rises monotonically in this wavelength region, which
means the temperature and the extinction are not easily separated.

For these reasons, the temperature of KN Cen B is determined using the
{\it IUE} Spectral Atlas of Morgan Keenan stars \citep{1983NIUEN..22Q....W}
The spectra used  are for B6 V ($\beta$ Sex), B3 V (17 Vul), and
B1 V (HD 31726).  As in \cite{2025ApJ...986...96E} the effect of the companion
on the observed (combined) colors was estimated from the calibration
of colors and absolute magnitudes from \cite{2000asqu.book..381D}.
For the Cepheid, M$_V$ =  -5.74 from \cite{2023A&A...672A..85C}.
The effects of the companion on the observed colors are summarized
in Table~\ref{com.eff} showing the observed colors, and the breakdown
into the Cepheid and the companion colors, as well as the derived E(B-V).  

\begin{deluxetable*}{lllllllllll}
\tabletypesize{\footnotesize}
\tablecaption{Effects of Companions on KN Cen\label{com.eff}}
\tablewidth{0pt}
\tablehead{
%\colhead{JD} & \colhead{Separation} & \colhead{PA} \\ 
  %\colhead{} & \colhead{$\arcsec$} & \colhead{$^o$}
\colhead{} &  \colhead{} & \colhead{Obs} & \colhead{} & \colhead{} & \colhead{Cep} & \colhead{} & \colhead{} & \colhead{Comp} & \colhead{} & \colhead{}  \\
\colhead{Spect.} & \colhead{B}  & \colhead{V} & \colhead{I$_C$} &\colhead{B}  & \colhead{V} & \colhead{I$_C$} & \colhead{B} & \colhead{V} & \colhead{I$_C$} & \colhead{E(B-V)} \\
  \colhead{Type} & \colhead{mag}  & \colhead{mag} & \colhead{mag} & \colhead{mag} & \colhead{mag} & \colhead{mag} & \colhead{mag} & \colhead{mag} & \colhead{mag} & \colhead{mag}\\
%\colhead{Type} & \colhead{} & \colhead{}  
% &  \colhead{}  & \colhead{mag} \\
}
\startdata
   B6 V &       11.477 & 9.855 &  7.989 &  11.50 & 9.87 &  7.993 &  15.27 & 14.72 & 14.004 & 0.787 \\   
   B3 V &       11.477 & 9.855 & 7.989  & 11.58 & 9.890 & 8.00  &  14.11 & 13.60 & 12.93 & 0.754 \\
   B1 V &      11.477 &  9.855 & 7.989 &  11.82 & 9.957 & 8.02  &  12.89 & 12.48 & 11.95 & 0.669 \\    
\enddata
\end{deluxetable*}

\begin{figure}
\includegraphics[width=6 cm,angle=90]{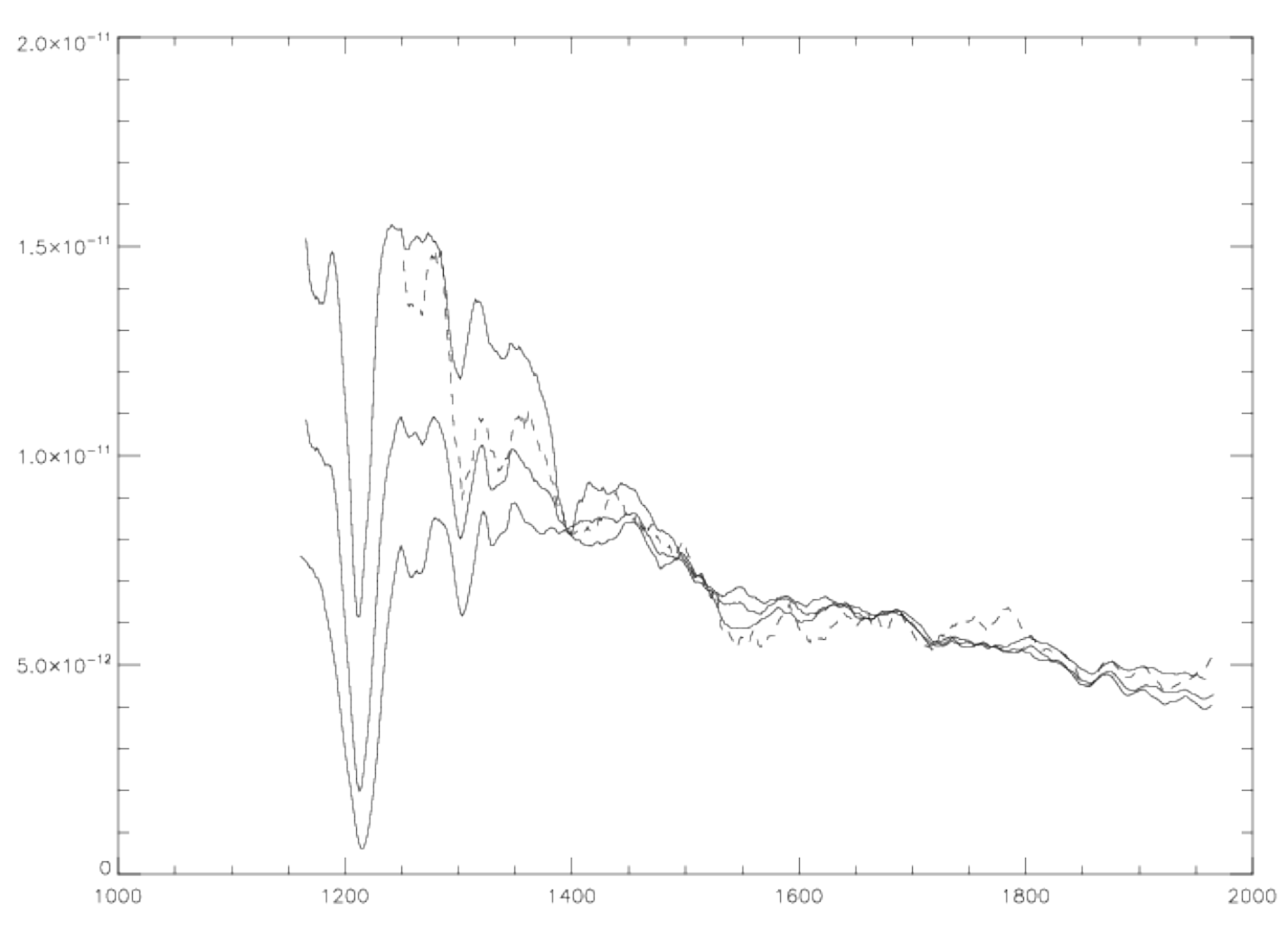}
%  \plotone{kncen_sum_cr_ebv69_sm10_srt.pdf}
  \caption{The spectrum of KN Cen compared with MK standard star spectra.
    The dashed line is KN Cen; the solid lines  (in descending order at
    1250 \AA\/ are B1 V, B3 V and B6 V. 
 Wavelength is in \AA; flux is in ergs sec$^{-1}$ AA$^{-1}$.  
  \label{kncen.comp}}
\end{figure}

In Fig~\ref{kncen.comp}, we show the summed spectrum of KN Cen compared with
B1 V, B3 V, and B6 V standard star spectra. The KN Cen spectrum is
clearly in the closest agreement with the B1 V star at 1250 \AA. 
The KN Cen spectrum in Fig.~\ref{kncen.comp}
is dereddened using E(B-V) = 0.69 mag.  Changing E(B-V) by a few 0.01's mag
does not alter this standard star selection.
A B1 V star has an effective temperature  of 26000 K from \cite{2013ApJS..208....9P} or 25500 K
from Drilling and Landolt.  This temperature implies that  KN Cen B is younger and hence
closer to the ZAMS, as shown in Fig.~\ref{fig:Tracks} and \ref{fig:Isochrone}, rather than being significantly cooler than an
isochrone for the Cepheid.  The temperature of KN Cen B may be perhaps a little cooler 
based on Fig~\ref{kncen.comp}:   closer to B3 V which would
have a temperature of 17000 K \citep{2013ApJS..208....9P}.  However, 
temperature measurements are not very precise in this range, and the resulting
companion is still closer to the ZAMS than the determination from
\cite{Evans.1994ApJ...436..273E}.

As a result, we conclude that the Cepheid KN Cen is \emph{not} a likely candidate to be a merger product.  The difference is the effective temperature from previous estimates and this is consistent with the resolution of the age discrepancy from the previous section and highlights some of the challenges for finding Cepheids that might have evolved from merger products. However, in the discussions of Sections 3 and 5 we continue to include the
previous value of the companion temperature as an illustration of this
approach. 

\subsection{SV Per and RW Cam} 

For SV Per, RW Cam the temperature of the companion was made as described in  \cite{2025ApJ...986...96E}.  
For these stars the analysis in \cite{Evans.1994ApJ...436..273E} concluded that the companions
were more luminous than ZAMS stars (appropriate to stars at the Terminal Age
Main Sequence [TAMS], and the spectral type of the companion was determined
from comparison with Class III stars in the IUE Spectral Atlas \citep{1983NIUEN..22Q....W}.  We adopt spectral types B8.0 III and B8.2 III for SV Per and RW Cam
respectively.

\section{Methodology} \label{sec:method}

Using ultraviolet IUE observations, \citet{Vitense.1985ApJ...296..175B,1992ApJ...384..220E} provided evidence for B-type companions for a number of previously known Milky Way Cepheids. In the visual bands, the B-type stars appear significantly fainter than the Cepheids, but due to their higher effective temperatures ($T_{\rm eff}$), they dominate the ultraviolet spectrum. \citet{Evans.1994ApJ...436..273E} derived temperature estimates and magnitude differences for a number of such systems, including the three discussed in this study. We adopt the companion properties reported in their analysis,  but have revisited them in Section 2.

The first step in this work is to analyze the extent of the age discrepancies. We compute stellar evolution tracks using Modules for Experiments in Stellar Astrophysics (MESA\footnote{\url{http://mesa.sourceforge.net}}; version 23.05.1), an open-source one-dimensional stellar evolution code based on adaptive mesh refinement \citep{MESA_2011ApJS..192....3P,MESA.2013ApJS..208....4P,MESA.2015ApJS..220...15P,MESA.Paxton2018ApJS..234...34P,MESA_2019ApJS..243...10P,MESA.2023ApJS..265...15J}. Our models assume a solar metallicity of $Z  = 0.02$, and a standard overshooting scheme with values prescribed by the 5M\textunderscore Cepheid\textunderscore blue\textunderscore loop test suite in MESA \citep{MESA.Paxton2018ApJS..234...34P}. For completeness, we utilize three overshooting values corresponding to standard, moderate and peak overshooting.  We do not include any exotic physics such as pulsation driven mass loss. We compute stellar evolution tracks ranging from $3$-$13$~ $M_\sun$ in steps of 0.25~$M_\sun$ and plot them on the HR diagram alongside the observed properties of the stars in our sample. We use these comparisons to estimate the masses and corresponding ages of the three B-type stars and the three Cepheids.

\subsection{Estimating the masses of the stars} \label{sec:mass}

\begin{figure*}[htbp]
       \centering
       \resizebox{\hsize}{!}{\includegraphics{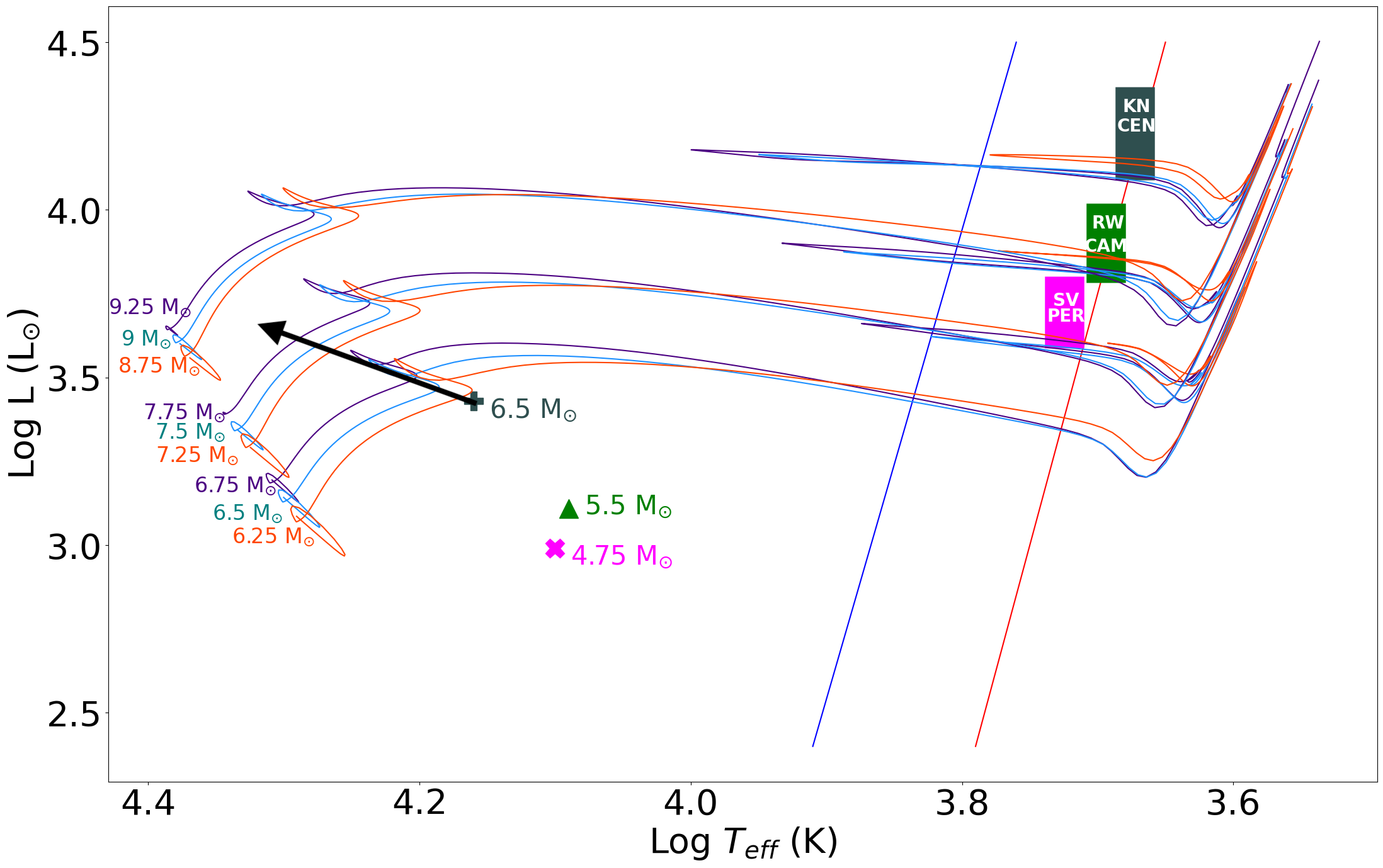}}
       \caption{Stellar evolution tracks to identify the three Cepheid masses in the blue loop\textsuperscript{\ref{fn:first}}. The purple, light-blue, and dark orange tracks correspond to standard, moderate, and peak overshooting schemes respectively. Assuming standard overshooting, the B-type companion mass estimates are annotated next to their corresponding markers. The arrow shows the approximate change of effective temperature and luminosity for KN Cen B.}
       \label{fig:Tracks}
\end{figure*}

\begin{figure}[htbp]
       \centering
       \resizebox{\hsize}{!}{\includegraphics{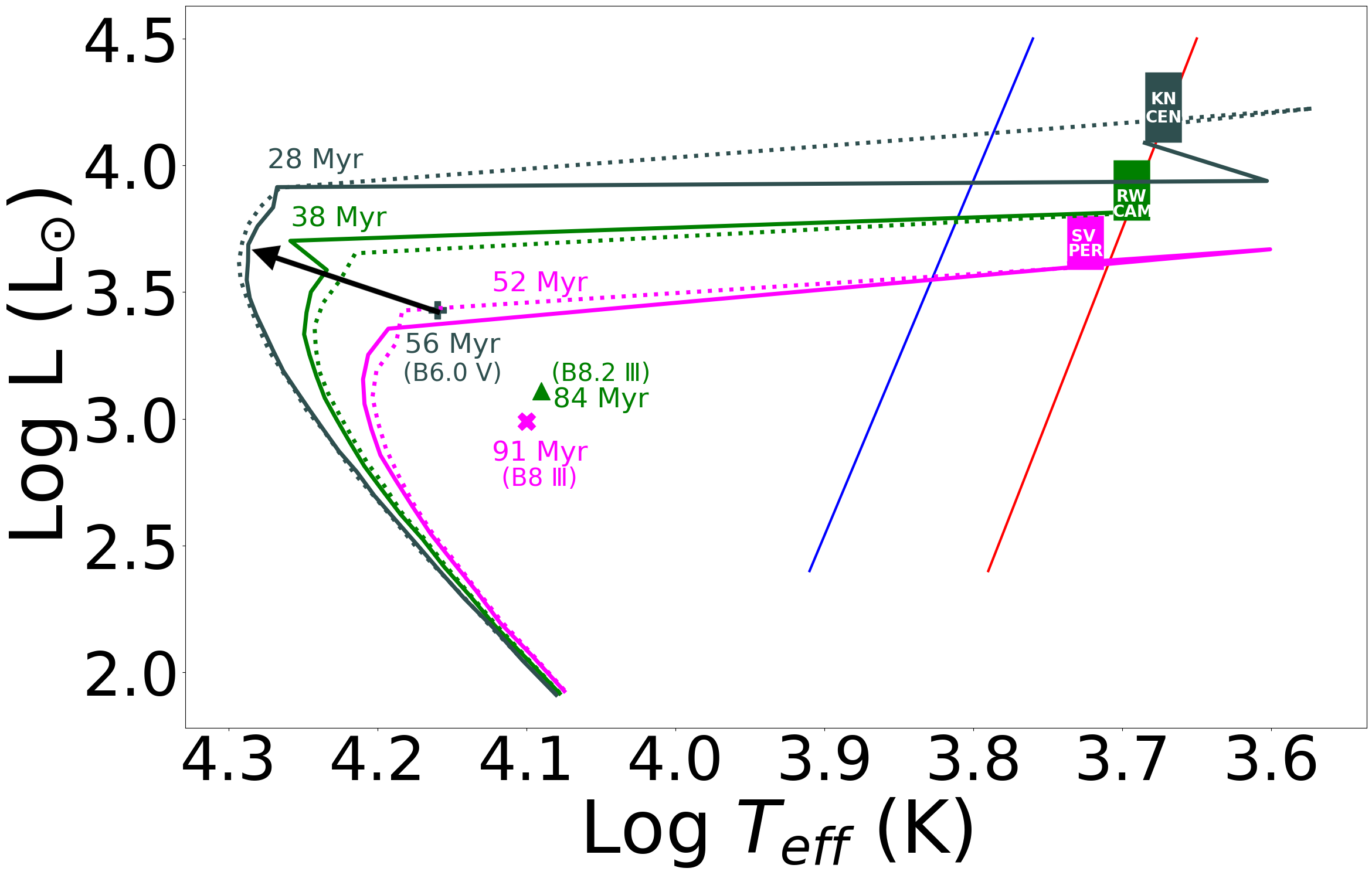}}
       \caption{Cepheid\textsuperscript{\ref{fn:first}} isochrones are colour coded, and demonstrate the apparent age discrepancy. The solid lines represent standard overshooting models, while the dotted lines indicate models with moderate overshooting. The arrow denotes the approximate difference in properties of KN Cen B based on the reanalysis of UV observations.}
       \label{fig:Isochrone}
\end{figure}

We show the evolutionary tracks for the three Cepheids in \autoref{fig:Tracks} along with the effective temperatures and luminosities  of the six stars. We note that decreasing the mass by roughly 0.25$~M_\odot$ and increasing overshooting results in nearly the same luminosity as the reference track, but with differences in the extent of the blue loop and the star's age. As a consequence, for SV Per, the blue loop of the track with the greatest overshooting value does not cross the red edge of the instability strip.\footnote{In Figures \ref{fig:Tracks}, \ref{fig:Isochrone}, \& \ref{fig:Mass Accretion} \label{fn:first}, the instability strip edges are plotted as established in the 5M\textunderscore Cepheid\textunderscore blue\textunderscore loop test suite in MESA} For this reason, the isochrone for the peak overshooting value is excluded from further analysis.

The predicted ages for stars evolving the blue loop, corresponding to the standard, moderate, and peak overshooting schemes, are tabulated in \autoref{tab:Summary Table} and the isochrones for the standard and moderate overshooting values are shown in \autoref{fig:Isochrone}. The first crossing for a Cepheid typically occurs on a timescale shorter by an order of magnitude compared to the blue loop phase \citep{Bono.2024A&ARv..32....4B}, accounting for only a small percentage of observed Cepheids \citep{Neilson.2012ApJ...760L..18N}. \citet{Berdnikov.2024AstBu..79..111B} suggested that KN Cen is on the third crossing based on a positive evolutionary period rate change $\dot{P}$ and the observed-calculated diagram. \citet{Berdnikov.2000ASPC..203..244B} tabulated a positive $\dot{P}$ for SV Per, and \citet{Turner.1998JAVSO..26..101T} estimated an increasing $\dot{P}$ for SV Per and RW Cam, advocating third crossing for both Cepheids based on the rate. The luminosity difference between the first crossing and the blue loop can be up to 0.30 dex \citep{Bono.2024A&ARv..32....4B}. Given the significant magnitude of the discrepancy, any such error in the age or luminosity of the Cepheid would be negligible. Therefore, the Cepheid is assumed to be in its blue loop phase which is consistent with period change observations and minimizes the age discrepancy. 

\begin{deluxetable*}{c |cc |cccc| cc }
\tablecaption{
The masses are calculated assuming standard overshooting. When overshooting is increased to its peak value, the lower mass limit decreases by up to 0.5 $M_\odot$, as illustrated in \autoref{fig:Tracks}. All uncertainties in the ages are within one million years (Myr). \label{tab:Summary Table}}
\tablehead{
\colhead{Star} & \multicolumn{2}{c}{Mass ($M_\odot$)} & \multicolumn{4}{c}{Age (Myr)} & \multicolumn{2}{c}{Age Discrepancy (\%
)}\\
\colhead{System} & \colhead{$M_{Cepheid}$} & \colhead{$M_{Companion}$} & \colhead{Standard} & \colhead{Moderate} & \colhead{Peak}  & \colhead{Companion} & \colhead{Minimum} & \colhead{Maximum}
}
\startdata
SV Per & $7.15^{+0.35}_{-0.4}$ & 4.75 & 52 & 58 & 65 & 84 & 29.2 & 61.5\\
RW Cam & $8.2^{+0.55}_{-0.45}$ & 5.5 & 38 & 43 & 48 & 91 & 89.5 & 139.4\\
KN Cen & $10.0^{+1.0}_{-0.75}$ & 6.5 & 28 & 30 & 30 & 56 & 86.6 & 100.0\\
\enddata
\end{deluxetable*}
In \autoref{fig:Isochrone}, we illustrate that the three Cepheids and their respective companions do \emph{not} fit the same isochrone. To compute the smallest age discrepancy, we select the isochrone age as the oldest possible age for the Cepheid, constrained by the mass-luminosity relation based on evolutionary predictions, which are strongly correlated with the PL relation. 
%%%%%%%%%%%%%%%%%%%%%%%%%%%%%%%%%%%%%%%

For stellar evolution models with  moderate values convective core overshooting, the masses and ages of the B-type companions are shown in \autoref{tab:Summary Table}. The prediction of luminosity and $T_{eff}$ of the companion lacks definitive error bars due to the limitations of measurements despite consistency with observations.

Thus, for comparison, we compute the required changes in effective temperature and luminosity for the companion to match the Cepheid within its measured uncertainties. This is illustrated on the HR diagram via the difference between the companion and the isochrone. On average, the companion must be at least 0.4 dex more luminous or 0.12 to 0.15 dex hotter to align with the isochrone. Assuming no changes to the luminosity of the companion, the required change in $T_{eff}$ corresponds to spectral type change from B8.III $\rightarrow$ B5.III for SV Per, B8.2III $\rightarrow$ B4.III for RW Cam, and B6.V $\rightarrow$ B2.V for KN Cen, using \citet{Zorec.2009A&A...501..297Z} for class III and \citet{Pecaut.2013ApJS..208....9P} for class V.

Moreover, evolutionary tracks indicate these companions are near or beyond the Terminal Age Main Sequence (TAMS). This observation is consistent with \citet{Vitense.1985ApJ...296..175B}, who noted that the companions have to be about 40\% hotter to be main sequence stars. At current temperature estimates, the companions have to be $>$ 0.75 $M_\sun$ smaller to be on the main sequence at a point earlier than the Sch$\ddot{o}$nberg-Chandrasekhar limit of the stars \citep{Chandrasekhar.1942ApJ....96..161S,Ziolkowski.2020MNRAS.499.4832Z}. A sufficiently greater luminosity would require the companion to have a relatively similar mass to the Cepheid; a scenario that has not been observed.

Our mass estimates agree with those of \citet{Vitense.1985ApJ...296..175B}, who used energy distribution with corrections for interstellar extinction along with stellar atmosphere models to determine the masses. For SV Per and RW Cam, assuming a metallicity of Y = 0.20, Z = 0.02, the authors estimate the Cepheid's mass depending on the crossing to range between 6.3-6.8 and 6.2-8 $M_\sun$ respectively. The masses of the respective companions are estimated to be 5.1 and 4.6 $M_\sun$. They also included a metallicity analysis for both the Cepheids at Z = 0.01 and 0.02, and Y = 0.20, 0.28, and 0.36, which immensely reduced the mass estimates \citep[See][Table 7]{Vitense.1985ApJ...296..175B}.  One issue to note is the dependence of the measured Cepheid masses on variations in metallicity. 

In our study, by accounting for higher metallicity in the PL relation and stellar evolution tracks, we find  that our best-fit masses might correspond to smaller luminosity estimates. SV Per and RW Cam are noted to have metallicity ranging from [Fe/H] = 0.01 - 0.07 and 0.04 - 0.11 \citep{Luck.2011AJ....142..136L, Lemasle2013A&A...558A..31L}, respectively, in accordance with their clusters which are moderately close to the solar metallicity. On the other hand, the metallicity of KN Cen is still debated, ranging from 0.07 \citep{Romaniello.2005A&A...429L..37R} to as high as 0.55 \citep{Genovali.2015A&A...580A..17G,daSilva2016A&A...586A.125D,Luck.2011AJ....142..136L}. A change in metallicity would consequently affect the PL relation \citep{Bhardwaj.2024A&A...683A.234B}; however, the extent of variation observed in the previous study is unlikely when evaluated through stellar evolution tracks, particularly in the case of younger stars such as Cepheids \citep{Bono.2000ApJ...543..955B,Takeda.2015MNRAS.450..397T,Toonen.2016ComAC...3....6T}.

\citet{Acharova.2012MNRAS.420.1590A} employed a period-mass relation \citep{Turner1996JRASC..90...82T} to estimate the masses of SV Per, RW Cam, and KN Cen to be 8.6, 10.4, and 15~$M_\sun$, respectively. 
These predictions are much greater than our best-fit results, likely due to the use of an older period-mass relation \citep{Turner1996JRASC..90...82T} that does not consider convective core overshooting.  Contrary to the relatively higher mass estimates , \citet{Kervella.2019A&A...623A.116K} measured the masses of SV Per and RW Cam to be $5.90^{+0.89}_{-0.89}$ and $6.70^{+1}_{-1} M_\sun$, respectively using a period-radius relation \citep{Gallenne.2017A&A...608A..18G,Caputo.2005ApJ...629.1021C}. Considering the lower limit of the stated masses, the corresponding ages of the Cepheids would predictably match the companion ages that we estimate. Using the Cepheid mass estimate and assuming a mass ratio of $q = 0.5$m based on the studies by \citet{Evans.2015AJ....150...13E}, the authors determined the masses of the companions to be close to 3 $M_\sun$. A 3 $M_\sun$ TAMS companion would be older than 400 Myr. This age would largely exceed the lifetime of a roughly 6 $M_\sun$ Cepheid. Milky Way Cepheids are ordinarily found in younger stellar populations ranging up to 300 Myr \citep{Martinez.2023MmSAI..94d..88M,DeSomma.2020MNRAS.496.5039D}, hence, the age discrepancy in SV Per and RW Cam persists.   

\citet{Anderson.2016ApJS..226...18A} estimated the age and mass of KN Cen in the blue loop phase to be 37 Myr and 8.6 $M_\sun$ based on their radial velocity studies for undetected companions. These estimates were based on a period-mass relation derived from the Geneva models and a period-age relation informed by rate of period change measurements. Despite a smaller mass and thus older age, the stated age does not align with an isochrone that includes a B6 companion. For consistency, the authors classified the companion as spectral type B2 and estimated the mass as 6.9 $M_\sun$ which matches the aforementioned required change in spectral type that we compute.

SV Per resides within the ASCC12\footnote{MWSC 427} cluster \citep{Medina.2021MNRAS.505.1342M} which is estimated to be over 260 Myr old \citep{Piskunov.2006A&A...445..545P,Maryeva.2021CoSka..51...78M,Kharchenko.2005A&A...440..403K}. Likewise, RW Cam is probably related to the Tombaugh 5 cluster \citep{Glushkova.2015BaltA..24..360G}, photometric studies of this cluster suggest its age to range between 185 and 250 Myr in most cases \citep{Baratella.2018AJ....156..244B,Zdanavi.2011BaltA..20....1Z,Zhang.2021A&A...654A..77Z,Romanyuk.2019AstBu..74..437R,     Jagadeesh.2021JApA...42..109J,Lata.2004BASI...32...59L}.

KN Cen is known to map out the Centaurus arm \citep{Majaess.2009MNRAS.398..263M,Majaess.2011ApJ...741L..27M}, but it is not clear which specific cluster it belongs to and the associated age. N-body simulation by \citet{Dinnbier.2024A&A...690A.385D} indicate that nearly 35\% of Cepheids escape to the field, KN Cen is suspected to be one such candidate (private communication).

These ages are much too old for either the Cepheid or the B-type companion. Nonetheless, small reductions in the luminosity or temperature estimates can greatly increase the age of the companions and can easily match those of the clusters. On the contrary, the stated cluster ages are longer than the lifetime of a Cepheid for any of the mass estimates discussed above. 

By comparing the representative or the lower limit of the masses from these studies  with our estimates from the standard overshooting tracks, we find the percentage difference for these Cepheids to be less than 26\% (Average:14\%) despite considering the most divergent values. Thus we conclude that the limitations of measurements and error bars cannot solely explain the incompatibility of each of the three binary star systems with their respective isochrones. Therefore, we concur with the previously proposed notion of these systems being age discrepant \citep{Evans.1994.2Page.IAUS..162...51E,Vitense1984IAUS..105..449B}.

\subsection{Mass Accretion Model} \label{sec: accretion}
To resolve the presented age discrepancy, we hypothesize that the three binary systems originally formed in triple star systems, likely hierarchical triplets, where two of the stars merged during main sequence evolution. However, when we reanalysis the UV spectra of KN Cen B, we find that the Cepheid KN Cen need not be a merger product.    As such, the Cepheid is not likely a rejuvenated merger product of the inner binaries.  The hypothesis is supported by population synthesis studies \citep{Hamers.2013MNRAS.430.2262H,Hamers.2021MNRAS.502.4479H} and is consistent with the fraction of progenitors found in binary or higher order systems. 
At least half of the known binaries with O- and B-type stars have been identified as triple systems \citep[][and references therein]{Toonen.2022A&A...661A..61T}. Even for low mass stars, the fraction of all B-type stars in triples is estimated to be about 50\% \citep[][and references therein]{Toonen.2016ComAC...3....6T}.
\citet{Evans.2005AJ....130..789E,Evans.2011BSRSL..80..663E} analyzed HST and IUE spectra for 18 binary Cepheids, revealing that 44\% of these have a third companion, bringing the fraction of Cepheids in triples to around 15\%.

Hierarchical triple systems are particularly intriguing because a tertiary companion induces angular momentum exchanges with the inner orbit, potentially leading to von Zeipel-Lidov-Kozai (hereby ZLK) oscillations \citep{Zeipel.1910AN....183..345V,Kozai.1962AJ.....67..591K,Lidov.1962P&SS....9..719L,Antognini2014MNRAS.439.1079A}. \citet{Toonen.2020A&A...640A..16T} studied stable hierarchical triplets with intermediate mass primaries showing that 70-80\% of these interact in their lifetime, with the primary star initiating mass transfer over 60\% of the time through a common envelope phase. A similar fraction has been indicated for high mass primaries \citep{Kummer.2023A&A...678A..60K}. \citet{Perets2012ApJ...760...99P, Hamers.2022ApJ...925..178H, Toonen.2022A&A...661A..61T} examined  triplets that are destabilized because of evolutionary factors and demonstrated that these instabilities can cause orbital disruptions, leading to collisions approximately 25\% of the time. Over two-thirds of these collisions involve the inner binaries, and almost all of them involve two main sequence stars. Therefore, common envelope evolution and collisions are fairly plausible resolutions for hierarchical triplets with stars on the main sequence and can lead to phenomena such as mergers, especially for sufficiently high eccentricity \citep{Perets.2009ApJ...697.1048P}.

Stellar mergers have been suggested to resolve various astrophysical anomalies \citep[see][for a review]{Henneco2024A&A...682A.169H}. A pertinent example is OGLE-LMC-CEP1812, a Cepheid estimated to be about 100 Myr younger than its red giant companion, for which a merger scenario is suggested to resolve the age discrepancy \citep{Pietrzynski.2011ApJ...742L..20P,Neilson.2015.OGLE-LMC-CEP1812A&A...581L...1N}. Other examples include HD 148937, HR 2949, and $\tau$ Scol, which exhibit large magnetic fields despite being early spectral-type high-mass stars \citep{Fossati2015A&A...582A..45F}, and accordingly have been identified as rejuvenated merger products. This is exemplified by age discrepancies with their companions \citep{Frost2024Sci...384..214F,Schneider2016MNRAS.457.2355S,Schneider2019Natur.574..211S}. Likewise, mergers address the apparent age spread in clusters such as Arches and Quintuplet due to the presence of stars such as blue stragglers \citep{Schneider2014ApJ...780..117S}. A similar age discrepancy has been suggested for starburst regions \citep{vanBever1998A&A...334...21V}.

We model the stellar merger as a mass accretion event applied to a single star model in MESA. Common envelope evolution is significantly more frequent than collisions with an impact parameter \citep{Toonen.2022A&A...661A..61T}. Thus, it is favourable to model this merger as a gentle accretion event.  \citet[See][Section 7]{MESA_2019ApJS..243...10P} details the treatment of the mass\textunderscore change function in a Lagrangian mesh. 

We evolve a star to a certain age $\tau$, then rapidly accrete mass to simulate a merger. Once the model accretes to the desired mass we continue evolution to the end of the blue loop stage. We assume composition of the accreted material has the same composition as the surface of the star itself. The accreted product is consistent with a Case A merger \citep{Gleebek.Merger.2008PhDT.......217G}.We note that a stellar merger does not conserve total mass and typically results in a loss of about 10\% to the interstellar medium  \citep{Gleebek.Merger.2008PhDT.......217G}.  We do not account for this mass loss in our models, but any difference is within the uncertainties of the measured properties of the stars in this work.  

We consider masses within the range $M_{\text{Cepheid}}/2 \leq M_{\text{Accretor}} \leq M_{\text{Companion}}$, in steps of 0.25 $M_{\sun}$. 
We evolve models of SV Per and RW Cam for roughly 75 Myr, and KN Cen for nearly 40 Myr, before accreting mass at a rate of $10^{-7} \sfrac{M_\sun}{\text{yr}}$ for SV Per and RW Cam, and $10^{-6} \frac{M_\sun}{\text{yr}}$ for the more massive KN Cen. The star's evolution continues until its luminosity aligns within the error bars of the Cepheid in the blue loop phase.

\section{Radial Velocities of Possible Merger Targets}

Since we are postulating that the Cepheids in the RW Cam and SV Per systems
are merger products,  they would not be expected to be in short period
binary systems.  This restriction does not apply to KN Cen, however.
Here we examine what velocity information is available about the binary/multiple
systems.

\subsection{RW Cam}

\begin{figure}
\includegraphics[width=9 cm,angle=0]{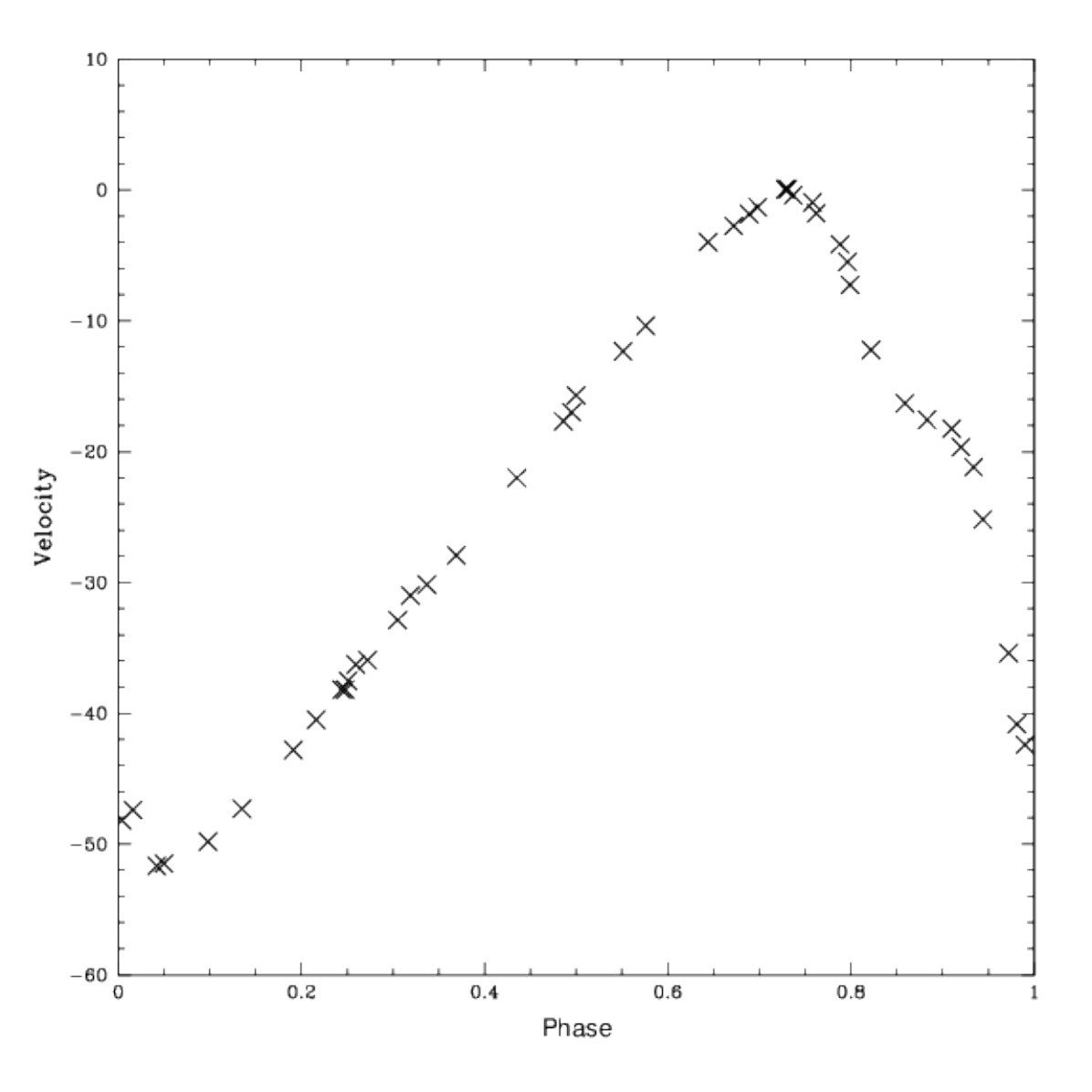}
%  \plotone{kncen_sum_cr_ebv69_sm10_srt.pdf}
\caption{The pulsation curve of RW Cam. Data are from VELOCE.)
  \label{rwcam.veloc}}
\end{figure}

Among high accuracy recent data for RW Cam, \cite{2024A&A...689A.224H} list those of \cite{2019A&A...631A..37B} and data in
preparation from Gorynya.  This discussion begins with the velocity date from VELOCE \citep{2024A&A...686A.177A}
shown in Fig,~\ref{rwcam.veloc}.  The pulsation period 16.414891 and epoch 2443840.949 from \cite{2022MNRAS.511.2125C} are used to plot the pulsation curve.  The data cover 7 years, and show very little variation due to
orbital motion.  \cite{2024A&A...690A.284S} find orbital motion in the dataset with an amplitude of 1.69 km s$^{-1}$.
Seven observations were made during the same period reported by \cite{2019A&A...631A..37B}, which agree with
 Fig,~\ref{rwcam.veloc} (using the velocities from centroid fitting for lines of all strengths).
As such, there is no indication  of a short period orbit (a few years) in the
 present system, but is likely a longer period orbit.

Using HST spatial scanning, \cite{2018ApJ...855..136R} found a companion within 0.2'' of the Cepheid.  At a distance of
 2400 pc (using the M$_V$ from \cite{2023A&A...672A..85C}), this is within 480 au, allowing plenty of
 space for a long period orbit.

Using proper motions from Gaia DR2, \cite{Kervella.2019A&A...623A.116K} find a strong proper motion anomaly for RW Cam   
compared with {\it Hipparcos} proper motions.  They estimate a maximum semimajor axis of 0.2'', and a
maximum orbital period of 400 years.

In summary, the combination of the radial velocities, proper motions, and spatial scans is consistent with a
wide orbit.  The system could plausibly have begun as a triple system, for which the inner binary could now
have merged into a single star.

% Borgn:  all line strengths, centroid vrs

\subsection{SV Per}

\begin{figure}
\includegraphics[width=9 cm,angle=0]{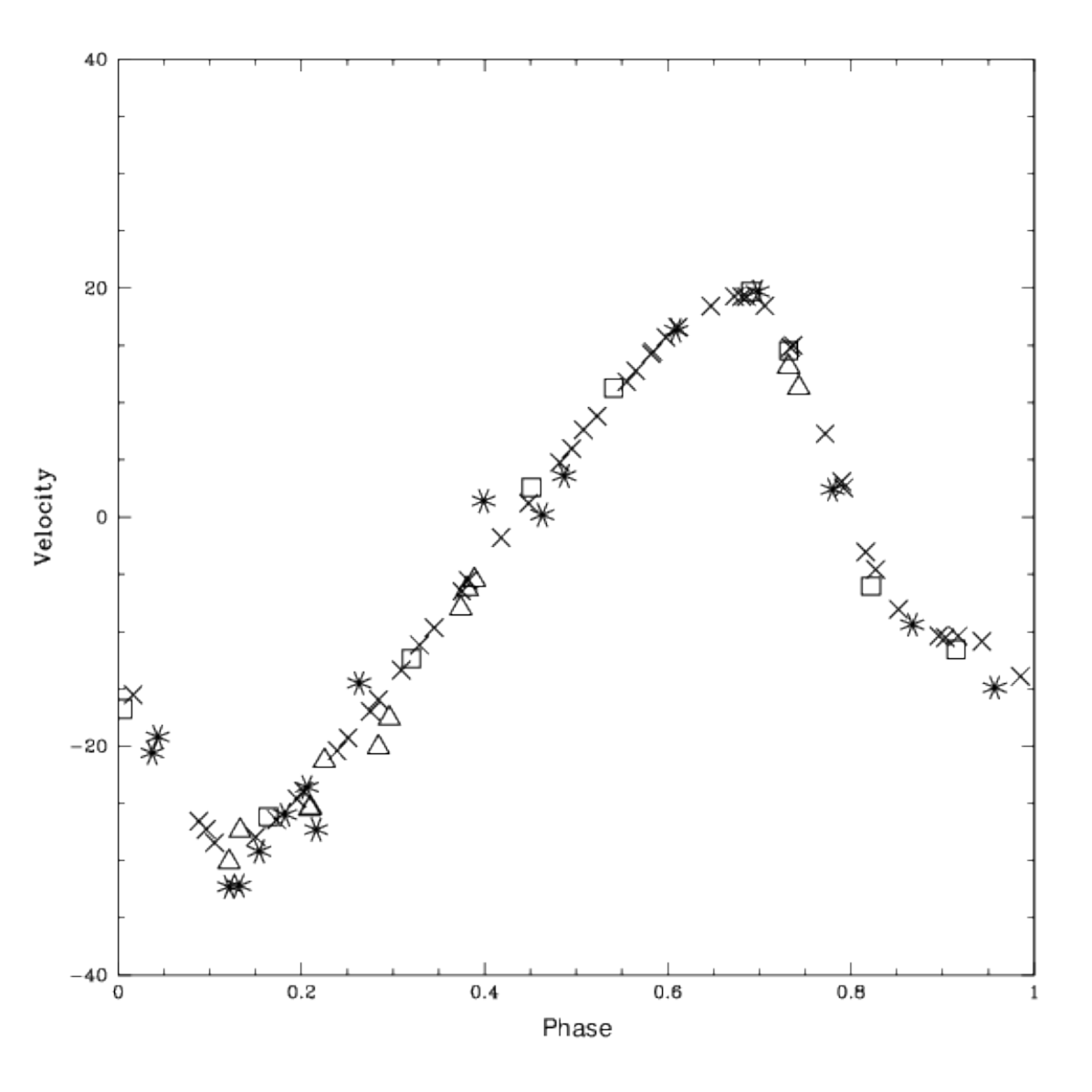}
%  \plotone{kncen_sum_cr_ebv69_sm10_srt.pdf}
\caption{The pulsation curve of SV Per. Symbols are: +: VELOCE; *: Harris;
  triangles: Gieren; and diamonds: Pont.  
  \label{svper.veloc}}
\end{figure}

The recent summary of high accuracy velocities by \cite{2024A&A...689A.224H} lists
\cite{1992A&A...253..126G}, Gorynya et al. (in preparation), and \cite{1997A&A...318..416P}
as having a substantial number of velocities. \cite{1992A&A...253..126G} discuss
previous velocity studies. In Fig.~\ref{svper.veloc} the pulsation data is
summarized from sources with the highest accuracy and significant number of
measurements, including velocities from Harris reported by \cite{1992A&A...253..126G}.
They present plots for all the data prior to 1992. The pulsation
period used is from \cite{2022MNRAS.511.2125C} P = 11.129319$^d$,
epoch T$_0$ = 2,443,839.303.  The VELOCE data provides new accuracy to the
pulsation curve and lengthens the data span (from the Harris data) to 42 years.
Fig.~\ref{svper.veloc} although there is some scatter to the data, no
orbital motion larger than 2-3 km s$^{-1}$ is seen, as was found by
Shetye, et al. (2024).

As with RW Cam, HST spatial scans \citep{2018ApJ...855..136R}
found a companion within 0.2'' of the Cepheid.
 At a distance of
3330 pc (using the M$_V$ from \cite{2023A&A...672A..85C}), this is within
670 au, allowing plenty of
 space for a long period orbit.
\cite{2019A&A...623A.117K} find a strong proper motion anomaly.
 Thus, in summary, velocities, spatial scans and the proper motion anomaly
 are all consistent with the hot companion in a wide orbit.

\subsection{KN Cen}
\begin{figure}
\includegraphics[width=9 cm,angle=0]{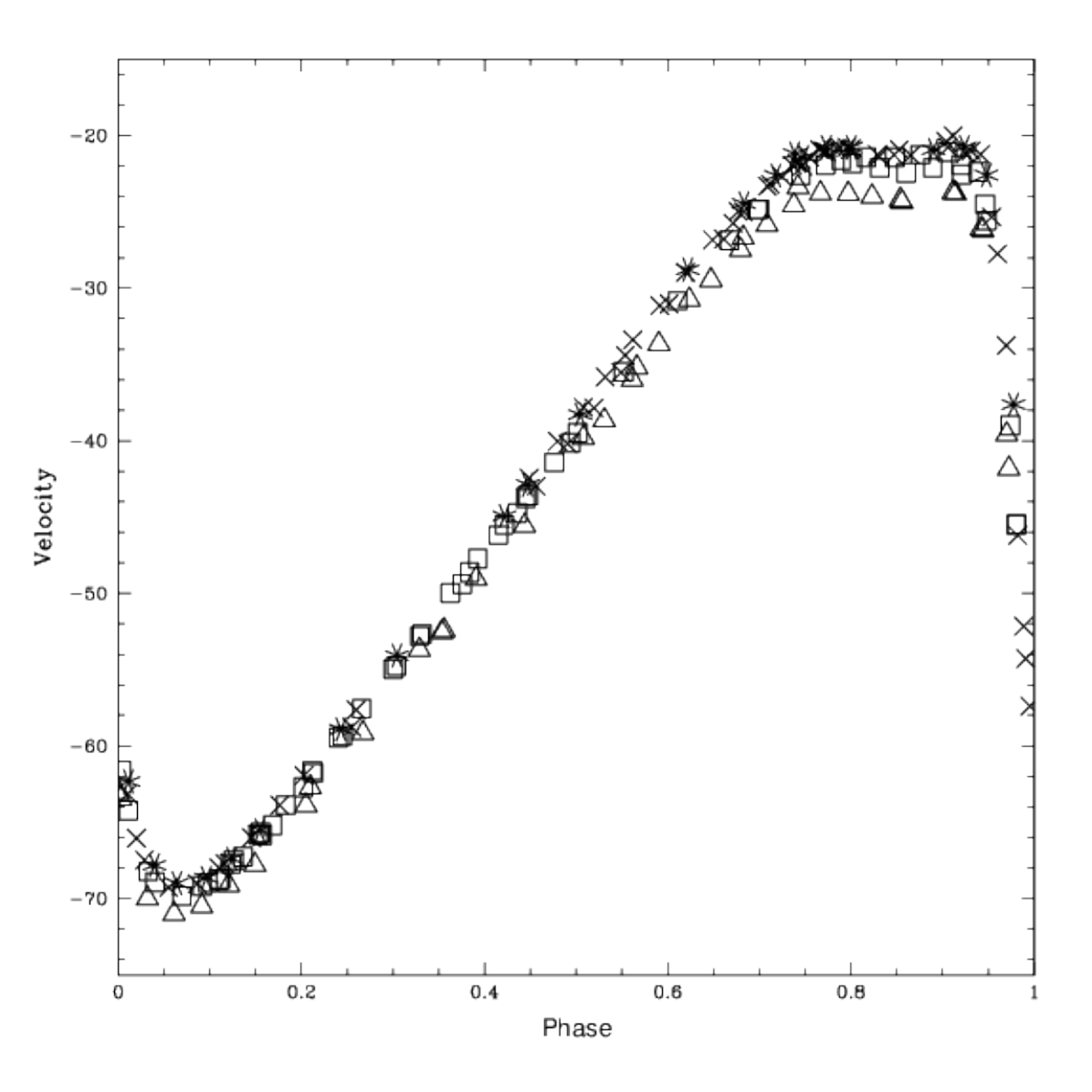}
%{ncen_veloc.pdf}
%kncen_veloc585_cc46}
%  \plotone{kncen_sum_cr_ebv69_sm10_srt.pdf}
\caption{The pulsation curve of KN Cen for four epochs of VELOCE data (see
  text for details).  Symbols are + for 565, *'s for 575, diamonds for 585, and
  triangles for 595.  
  \label{kncen.veloce}}
\end{figure}

          Since the Cepheid KN Cen does not appear to be a merger product,
the orbit of the Cepheid and the hot companion could be as short at a year.
 There are two extensive velocity datasets for KN Cen: VELOCE  \citep{2024A&A...686A.177A} and that
 of \cite{1985SAAOC...9....5C}. The original discussion of the data in the VELOCE sample \citep{Anderson.2016ApJS..226...18A}
was the first to identify orbital motion.
They find
possible orbital velocity variation of 3 km/s between datasets taken 10,000 days (27 years)
apart.  This is consistent with a long period orbit.
Fig.~\ref{kncen.veloce} shows the orbital motion in the VELOCE dataset
(Anderson et al, 2024).  The velocities in Fig.~\ref{kncen.veloce} are plotted
using the pulsation period from \cite{2022MNRAS.511.2125C}  P = 34.029641$^d$,
T$_0$ = 2,436,239.4202.  Data are shown divided into 4 BJD groups 2,456,783 to
2,456,818; 2,457,046 to 2,457,847; 2,458,151 to 2,458,620; and 2,459,318 to
2,459,639 referred to as 565, 575, 585, and 595 respectively in
Fig.~\ref{kncen.veloce}.  The orbital motion between these periods is
clearly shown. 
The discussion of binaries within the
    VELOCE dataset \citep{2024A&A...690A.284S} confirms the binary motion,
 however,  data was insufficient to determine the orbit.
 The VELOCE sample covered nearly
    8 years (with additional data from the 2016 sample), see
    Fig.~\ref{kncen.veloce}.

\begin{figure}
\includegraphics[width=9 cm,angle=0]{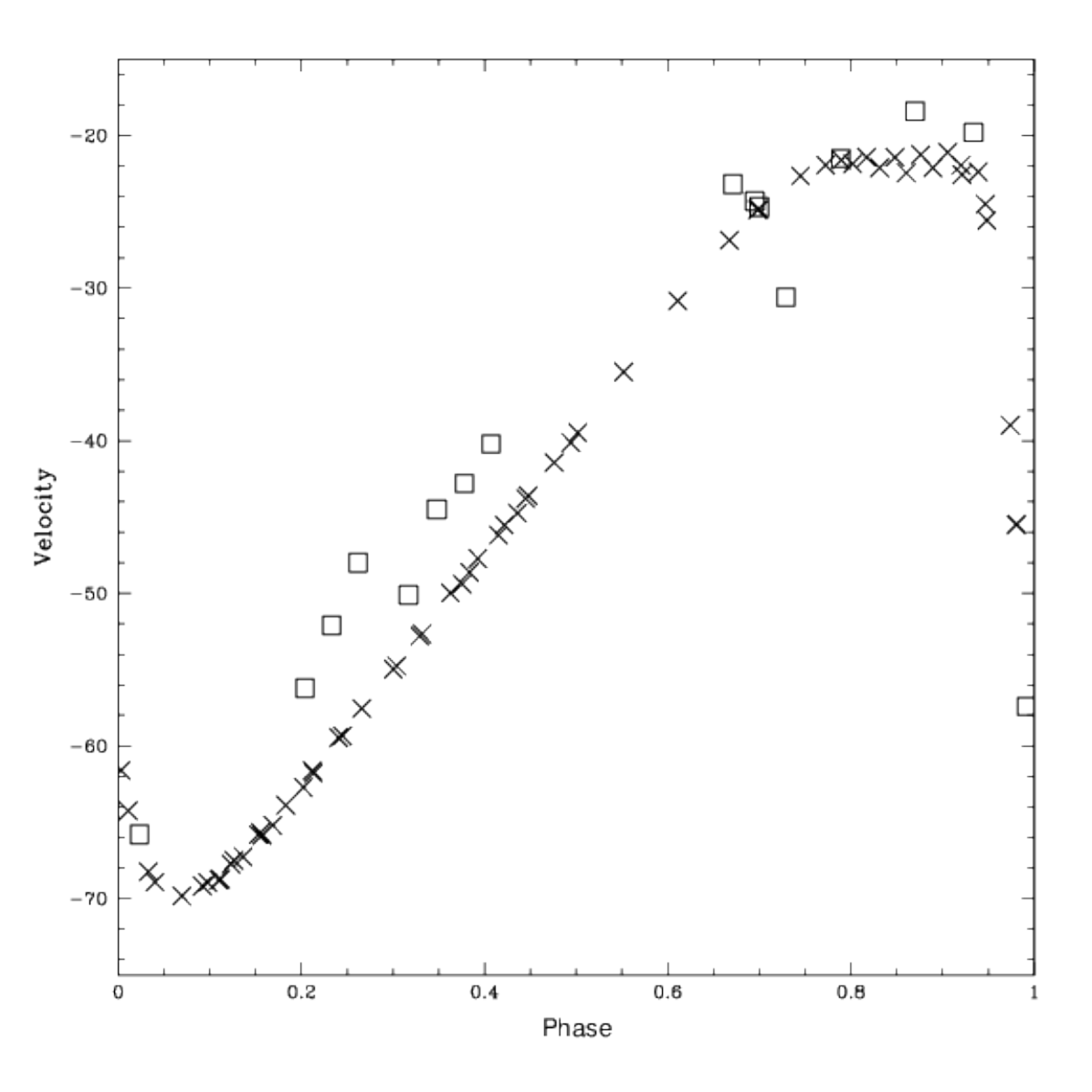}
%{kncen_veloc585_cc46.pdf}
%  \plotone{kncen_veloc585_cc46.pdf}
\caption{The pulsation curve of KN Cen for one epoch of VELOCE data
  compared with one year of velocities from\cite{1985SAAOC...9....5C}. Symbols:
  VELOCE data 585: +'s; Coulson and Caldwell data 46: diamonds.
  \label{kncen.vrcc}}
\end{figure}

         The earlier data from Coulson and Caldwell are less accurate than the
      VELOCE velocities which have a typical accuracy of less
      than 0.1  km s$^{-1}$.  The Coulson and Caldwell velocities have an
      estimated accuracy of $\pm$ 2.5  km s$^{-1}$
      \citep{1985ApJS...57..595C}.
      They have a duration of 3 years.
      Fig.~\ref{kncen.vrcc} shows one of the more numerous VELOCE groups (585)
      compared with one of the more numerous years from Coulson and Caldwell
      (JD 2,444,653 to 2,444,778 called 46).  To make this comparison, an
      estimated phase shift of 0.05 was added to the Coulson and Caldwell
      phases.  One reason the phase shift is only approximate is that  the
      O minus C diagram of \cite{2022MNRAS.511.2125C} exhibits both an evolutionary
      parabola and a wave.  The origin of the wave is unknown but it is
      frequently found in long period Cepheids such as KN Cen.
        The combination of the two
          datasets is not adequate to determine an orbit.  However orbital
          motion has not shifted velocity curve by much nor has  the
          spread from orbital
          motion increased.  
          This suggests that the
          orbital velocity amplitude is small and confirming the
          period range of a few years. Proper Motion anomaly from {\it Gaia} DR2 is suspected \citep{Kervella.2019A&A...623A.116K}.

\section{Results \& Discussion} \label{Sec: Discussion}

 For SV Per and RW Cam we find an age discrepancy between the Cepheid and
the companion.  For KN Cen, the new data and analysis are consistent with a
hotter companion and a single isochrone.  

To resolve the age discrepancy, we outline four potential resolutions based on the preceding analysis as follows: 1) hotter B-type companions, 2) overestimated Cepheid mass, 3) stellar capture, and 4) stellar mergers. We have shown that for a hotter companion to be consistent with observations, the spectral types of the companion stars would need to be systematically wrong.  \cite{Keller.2006ApJ...642..834K, Keller.2008ApJ...677..483K} demonstrated that, on average, Cepheid masses computed using stellar evolution models tend to be about 10\% greater than mass estimates from stellar pulsation models. Even if we assume that the Cepheids in this study are 20\% less massive, a significant age discrepancy remains. Another possibility is that the Cepheids and their companions may not have evolved coevally, and instead became gravitationally bound in a stellar encounter. \citet{Bailer-Jones.2018A&A...616A..37B,Bailer-Jones.2018A&A...609A...8B,Bailer-Jones.2015A&A...575A..35B} estimated the stellar encounter rate $\Gamma$ within 5 parsecs of the Sun to be approximately 0.0005 yr$^{-1}$, based on Gaia and Hipparcos data. For comparison, the rate of head-on collisions in Galactic destabilized triplets is estimated to be similar, on the order of $10^{-4}$ yr$^{-1}$ \citep{Perets2012ApJ...760...99P}, both of which are significantly lower than the inferred rate of all Galactic mergers of about 0.5 yr$^{-1}$ \citep{Kochanek.2014MNRAS.443.1319K}. 

\begin{figure}[htbp]
    \centering

%    \begin{subfigure}{\linewidth}
        \centering
        \includegraphics[width=0.85\linewidth]{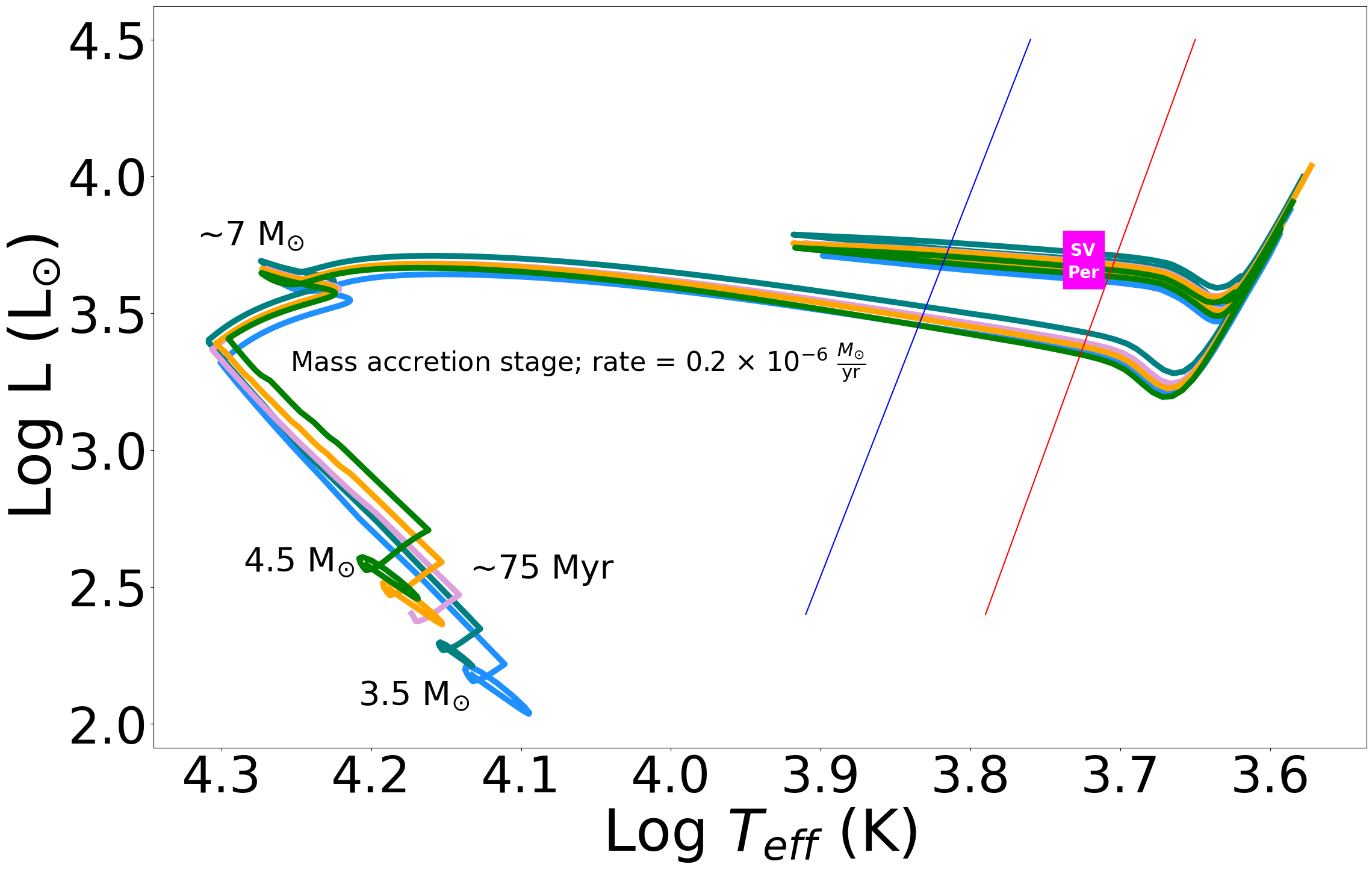}
 %       \caption{SV Per mass accretion model.}
%        \label{fig:SVPer}
%    \end{subfigure}
    \vspace{1em}

  %  \begin{subfigure}{\linewidth}
        \centering
        \includegraphics[width=0.85\linewidth]{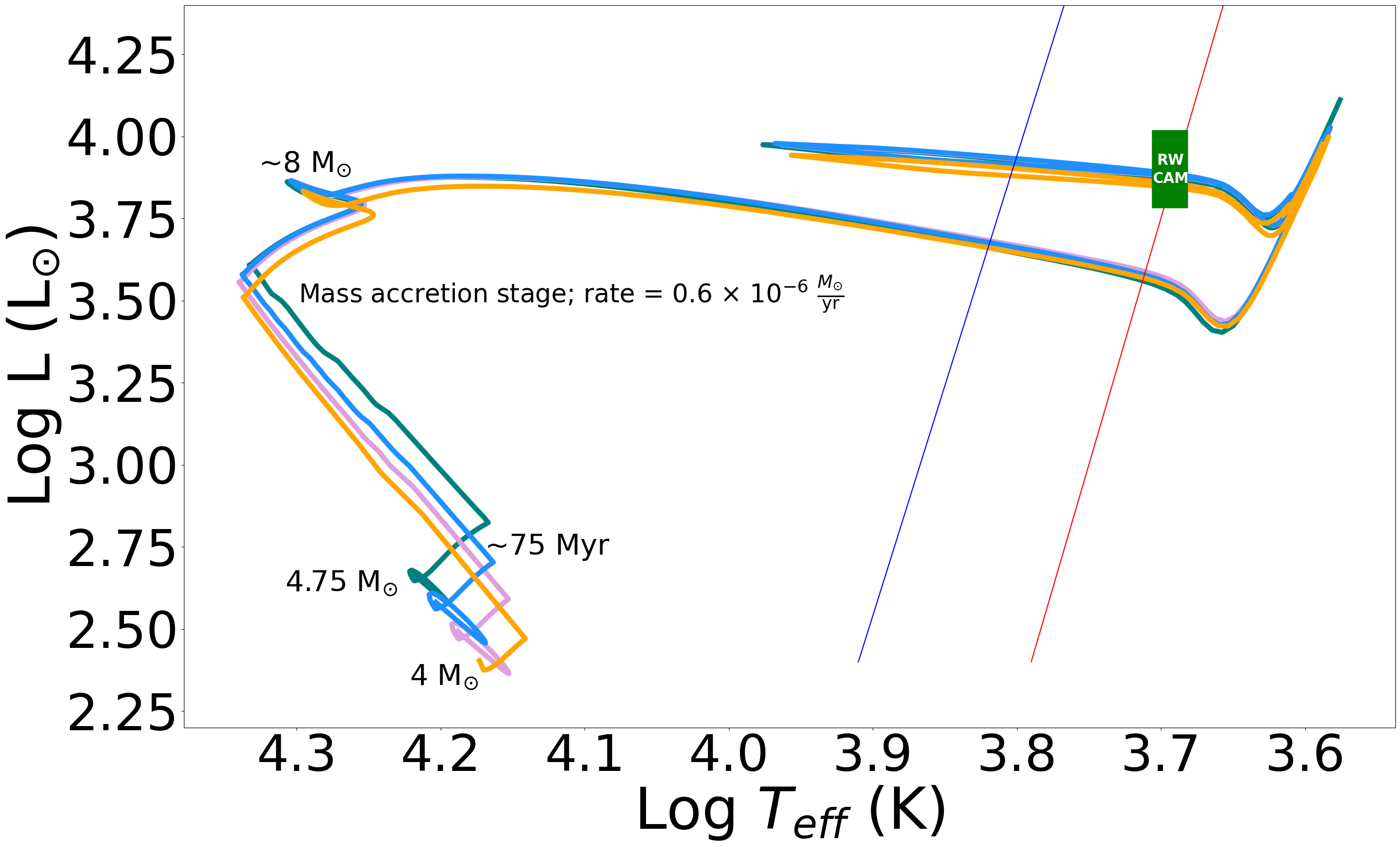}
    %    \caption{RW Cam mass accretion model.}
      %  \label{fig:RWCam}
    %\end{subfigure}
    \vspace{1em}

    %\begin{subfigure}{\linewidth}
        \centering
        \includegraphics[width=0.85\linewidth]{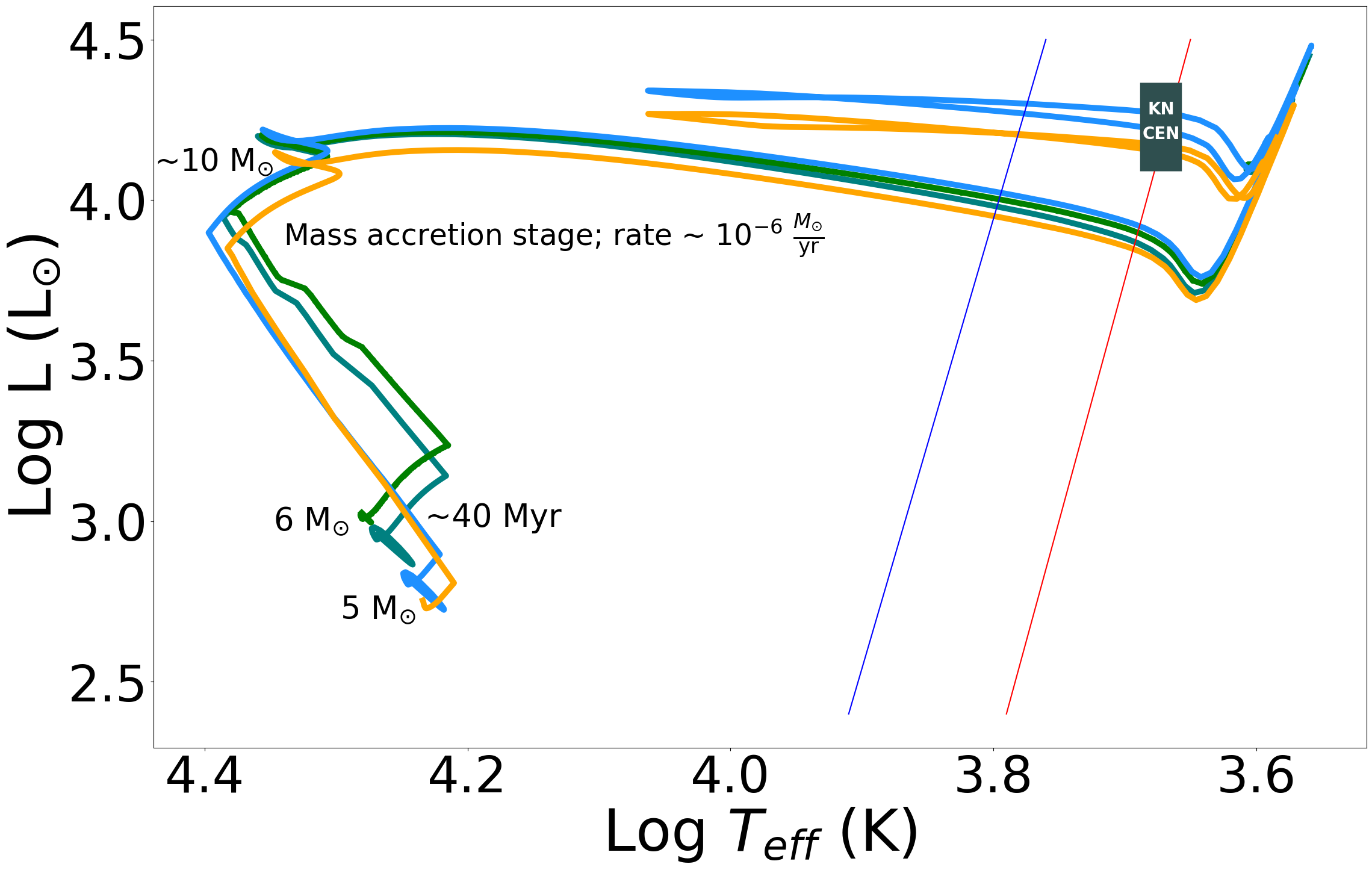}
     %   \caption{KN Cen mass accretion model.}
      %  \label{fig:KNCen}
    %\end{subfigure}

    \caption{Stellar evolution tracks\textsuperscript{\ref{fn:first}} for the mass accretion models of SV Per (top), RW Cam (middle), and KN Cen (bottom). The blue and red lines mark the approximate boundaries of the Cepheid instability strip. Each panel shows consistency with the observed luminosity and temperature from previous measuresments, and an age match with the corresponding B-type companion.}
    \label{fig:Mass Accretion}
\end{figure}

The most plausible resolution of the four is that each system was born in a hierarchical triplet and the Cepheid is the rejuvenated merger product of two main sequence stars that has since evolved and thereby appears younger. The rejuvenated stellar evolution tracks in \autoref{fig:Mass Accretion} demonstrate that a Cepheid resultant from a merger would have the same luminosity and $T_{\rm{eff}}$ as a lone Cepheid observed today. Moreover, the age of the merged Cepheid would align with its companion and the cluster, resolving the age discrepancy.KN Cen is also shown in Fig.~\ref{fig:Mass Accretion}, although it is no longer considered a merger product.

Simplified mass accretion models like those computed  here, lack potentially important physics such as  magnetic fields, internal chemical gradients and surface compositions \citep{Henneco2024arXiv240614416H}. Regarding chemical composition, main sequence mergers are expected to have greater luminosities than single stars of the same mass due to an increase in mean molecular weight from helium mixing in the interior part of the star's outer envelope, as observed in blue stragglers \citep{Glebbeek.2008IAUS..246..363G}. It is unclear whether this enhanced luminosity is observable in the three Cepheids. \citet{Rui.2021MNRAS.508.1618R} simulated a main sequence primary nearing TAMS that merges with another main sequence star and suggest that such a merger product, despite having depleted much of its hydrogen core, would be indistinguishable from a lone star of the same mass. \citet[][see their figure 12]{Dinnbier.2024A&A...690A.385D} found that approximately 65\% of mergers pertaining to Cepheids will occur during the envelope expansion close to TAMS, similar to the previously described scenario. Therefore, akin to our models, significant internal and surface chemical composition changes in the Cepheids, despite a merger, are less likely to be observed, especially given that the three Cepheids are reasonably massive and ZLK oscillations likely led to a merger comparatively earlier to the norm in their lifespan.

\citet{Barron.2022MNRAS.512.4021B,Barron.2022arXiv220709255B} observed 15 Cepheids and detected Stokes V parameters for over 50\% of them, including confirmation of the previously detected magnetic field in $\eta$ Aql \citep{Butkovskaya.2022AcAT....3a...1B}. Their results imply that Cepheids typically have weak surface magnetic fields ($<$1G) \citep{Wade.2024BSRSL..93..514W}, likely harboured by their convective atmospheres with no indication of strong magnetic fields ($>$1kG) that might be consistent with stellar mergers \citep{Langer.2012ARA&A..50..107L}.  

Despite the minimal limitations of fast accretion as discussed in this context, an alternative approach would involve using a dedicated merger code and then integrating the resulting structure into an evolutionary model. For instance, \citet{Shiber.2024ApJ...962..168S} tested the hypothesis of Betelgeuse as a merger product using the code Octo-Tiger \citep{OCTO-TIGER.2021MNRAS.504.5345M}, followed by stellar evolution modelling with MESA. Such a transition requires approximations and well-constrained parameters, such as radius and mass, which are known with a certain degree of uncertainty for Betelgeuse, but less accurately for the three Cepheids. Therefore, accretion remains a suitable and fairly accurate method for this kind of modelling.

To summarize, we investigate and confirm an apparent age discrepancy for two of the three Galactic Cepheid binaries using mass estimates from stellar evolution tracks in MESA. Using an updated temperature measurement where the B-type companion is significantly hotter than previously measured, we find that KN Cen is consistent with not being a merger.  The results demonstrate that the two Cepheids have to be at least 1.5 $\times$ older to be coeval with the companions. To resolve this, we stipulate four resolutions in the forms: hotter B-type companions, Cepheid mass discrepancy, stellar capture, and stellar mergers, discussing the required extent and comparing them with literature values. We focus on the merger theory, hypothesizing that these binaries formed in hierarchical triple star systems, with the Cepheids evolving from the coalescence of the inner binaries, presumably as a result of ZLK oscillations. It may be possible that the two progenitors formed without the third star having any significant gravitational impact on the system, but it would require an even more specific initial configuration.

Further, we test the consistency of stellar mergers that are simulated as a mass accretion event applied to a single star model in MESA. We evolve a star to a certain age $\tau$, rapidly accrete mass to imitate a merger, and then continue evolution of the accreted star to the end of blue loop phase. The simulation demonstrates that the Cepheids are consistent with merger products in fits of effective temperature and luminosity, along with suitable isochrones. Thereby making stellar merger the most probable and plausible resolution among those stipulated. The Jupyter notebooks used for analysis are available on  \href{https://github.com/Sashwat-Sashi/Cepheid\_Age\_Discrepancy.git}{https://github.com/Sashwat-Sashi/Cepheid\_Age\_Discrepancy.git}.

\begin{acknowledgements}
The analysis has used NumPy \citep{Numpy.2020Natur.585..357H} and Matplotlib \citep{Matplotlib.2007CSE.....9...90H}, for which we express gratitude to the developers. The preparation of this article has made use of the Astrophysics Data System, funded by NASA under Cooperative Agreement 80NSSC21M00561, and the ASTRO-PH.SR e-print server.  HRN acknowledges funding from the Memorial University of Newfoundland and from the National Science and Engineering Research Council.
\end{acknowledgements}

\bibliography{bib.bib}
\bibliographystyle{aasjournal.bst}
\end{document}